\documentclass[10pt]{article}    

\usepackage{cite} 
\usepackage{url}  
\usepackage{ifthen}  
\usepackage{multicol}   
\usepackage{listings}
\usepackage{comment}
\usepackage{rotating}
\usepackage[utf8]{inputenc} 
\usepackage{amsmath,amssymb,graphics,graphicx} 
\usepackage{color}

\newboolean{publ}

\makeatletter

\newcommand{\fig}[1]{Fig. \ref{#1}}

\makeatother

\begin{document}

\begin{center}
\noindent {\Large Prediction stability in a data-based, mechanistic model of $\sigma^F$ regulation  during sporulation in  \textit{Bacillus subtilis}}\\
\vspace{2mm}
{Georgios Fengos$^{1}$ and Dagmar Iber $^{1,*}$}
\end{center}
\vspace{10mm}

\noindent $^{1}$Department of Biosystems Science and Engineering (D-BSSE) Eidgen\"{o}ossische Technische Hochschule Zurich (ETHZ) and Swiss Institute of Bioinformatics (SIB), Mattenstrasse 26, 4058 Basel, Switzerland\\

 $^{*}$Correspondence to: dagmar.iber@bsse.ethz.ch


\section*{Abstract}

Mathematical modeling of biological networks can help to integrate a large body of information into a consistent framework, which can then be used to arrive at novel mechanistic insight and predictions. We have previously developed a detailed, mechanistic model for the regulation of $\sigma^F$ during sporulation in \textit{Bacillus subtilis}. The model was based on a wide range of quantitative data, and once fitted to the data, the model made predictions that could be confirmed in experiments. However, the analysis was based on a single optimal parameter set. We wondered whether the predictions of the model would be stable for all optimal parameter sets. To that end we conducted a global parameter screen within the physiological parameter ranges. The screening approach allowed us to identify sensitive and sloppy parameters, and highlighted further required datasets during the optimization. Eventually, all parameter sets that reproduced all available data predicted the physiological situation correctly. 
\newpage

\section*{Introduction}
 
Mathematical modeling is the method of choice to test hypotheses and make novel predictions concerning complex biological signaling networks \cite{Iber:2012hm}. Decades of biochemical experiments and recent advances in high-throughput technology provide us with an increasing amount of quantitative data on protein interactions. Once integrated in detailed mathematical models, mathematical models can be used to delineate and clarify regulatory mechanisms of protein signaling networks, and to make novel predictions of unexpected properties of the system, which can be then tested experimentally \cite{Iber:2012jp, steven2003computational}. The predictive power of these models depends on the accuracy of both the network and the parameter values. Experimental data is typically scarce, and many kinetic parameter values and cellular protein concentrations are either unknown or only possible to measure \textit{in vitro}. Cellular concentrations of signaling proteins  vary substantially across different cellular conditions, and estimated \textit{in vitro} values may not always reflect the \textit{in vivo} behavior \cite{minton2001influence}.

Model-based experimental design is used in many fields of science to optimize the planning of the necessary experiments \cite{mead1990design}. Similar strategies have also been explored in biology \cite{kreutz2009systems,cho2003experimental,faller2003simulation}, but the design strategies are typically heavily dependent on initial parameter choices, and given the experimental restrictions, there is often not much scope for further experimental design beyond what is possible by verbal reasoning. Often data is therefore acquired before modelling, and parameter values are then estimated using standard parameter identification procedures \cite{Jaqaman:2006p24311, Geier:2012gl, peifer2007parameter,voss2004nonlinear}. Local parameter optimization procedures typically yield one optimal parameter set together with confidence intervals. Global optimization procedures can yield a larger set of distinct parameter sets that all reproduce the data similarly well \cite{moles2003parameter,mendes1998non}. Sensitivity analysis is often  carried out to analyse how parameter perturbations around the global optimum/optima affect model predictions \cite{zi2011sensitivity,cho2003experimental,sahle2008new},

The experimental observations usually do not suffice to unambiguously determine all parameter values, and many parameter values remain poorly constrained even when detailed quantitative data is available because the measured model output is insensitive to these parameter values for the applied perturbations \cite{hengl2007data}. This issue has been recognized as ''parameter sloppiness" \cite{gutenkunst2007extracting}. If model predictions are based on one optimal parameter set then a parameter value must also be fixed for the sloppy parameters. If the model prediction of interest depended on one such sloppy parameter then model predictions may be severely hampered. One way to avoid such shortcomings is to use an ensemble approach. Ensemble approaches can be used to explore the type of model predictions that can be made for the sets of possible parameter values \cite{Melke:2006p5755, Kuepfer:2007p47476}. Clustering can then be used to classify the different predictions. 

We have previously studied a detailed model for the regulatory network that controls cell differentiation during sporulation in \textit{Bacillus subtilis} \cite{iber2006mechanism, Iber:2006ek}. The model was based on detailed, quantitative measurements and the parameter values were adjusted such that all data was reproduced very well. The model was subsequently used to explore the physiological situation. The model was sufficiently powerful to predict that the \textit{in vivo} rate of the key phosphatase SpoIIE had to be 100-150 lower than what had previously been measured in  \textit{in vitro}. This model prediction was confirmed in experiments, thus demonstrating the predictive power of the model. The model further revealed the allosteric nature of the interaction between SpoIIAA and SpoIIAB, which could also be confirmed in experiments. The model was subsequently used to  define the mechanism by which activation of the transcription factor $\sigma^F$ is achieved in the smaller pre-spore, but avoided in the larger mother cell upon asymmetric cell division during sporulation. The model predicted that this would be the result of the difference in cell size. Thus the phosphatase SpoIIE is a membrane protein and concentrates on both sides of the septum, while both its substrate SpoIIAA and the kinase, SpoIIAB, are cytoplasmic proteins. Given the smaller size of the pre-spore, the activity of SpoIIE is higher in this compartment, thus triggering the release of $\sigma^F$ from an inhibitory complex with SpoIIAB by dephosphorylating SpoIIAA; unphosphoryalted SpoIIAA binds to SpoIIAB and displaces bound $\sigma^F$. The model further showed that the small difference in SpoIIE activity is sufficient to result in differential cell fate because of the cooperative binding of SpoIIAB and SpoIIAA, and because of the low turn-over rate of the phosphatase \cite{iber2006mechanism, Iber:2006ek}. As a result of the low turn-over rate of SpoIIE, most SpoIIE is bound to phosphorylated SpoIIAA, and accumulation of SpoIIE on the septum therefore increases the concentration of SpoIIAA in the smaller prespore. The model also suggested that an important driving force behind the organization of  the genes \textit{spoIIAA}, \textit{spoIIAB}, and \textit{sigmaF}  on the \textit{spoIIA} operon may have been protection from molecular noise \cite{iber2006quantitative}. Addition of physiological noise levels  to the expression of these genes in the model lead to a reduction of the sporulation efficiency to 40\% if \textit{spoIIAB} was expresssed separately  and to 60\% if \textit{spoIIA} was expressed separately.   These predictions matched the observed sporulation efficiencies that were obtained when either \textit{spoIIAB} or \textit{spoIIA} were expressed separately, but under the same promotor. 

The above predictions were all made with  a single parameter set. While experiments confirmed the predictive power of the model, we wondered whether similar predictions would have been obtained for all physiological parameter sets that allow us to reproduce  the experimental data. We therefore conducted a global parameter screen within the physiological parameter ranges and restricted these with the quantitative data using evolutionary sampling \cite{beyer2002evolution}. Although exhaustive sampling of the physiological parameter space is impossible, qualitative model behaviours typically cluster within the parameter space, thus permitting an ensemble approach  \cite{geier2011computational,battogtokh2002ensemble,celliere2011plasticity,Melke:2006p5755, Kuepfer:2007p47476}. We subsequently tested in how far the parameter sets that allowed us to reproduce the measured \textit{in vitro} data would allow use to reproduce also the \textit{in vivo} behaviour. Initially we focused only on the published kinetic data. We found that the parameter sets, which allowed us to reproduce all measured time courses, still failed to reproduce the \textit{in vivo} behavior of the network. Comparison of the successful and unsuccessful simulations highlighted the critical parameters. Upon constraining these by further published data, all simulations predicted the physiological data correctly, even though many parameter values remained poorly restricted. We conclude that evolutionary sampling can provide a broader view on the qualitative behaviour of complex physiological networks, and it confirms that for the available data, the model exhibits a high stability of its predictions, once the key parameter values are constrained.

\section*{Results}

\begin{figure}[!ht]
\begin{center}
\includegraphics[width=8cm]{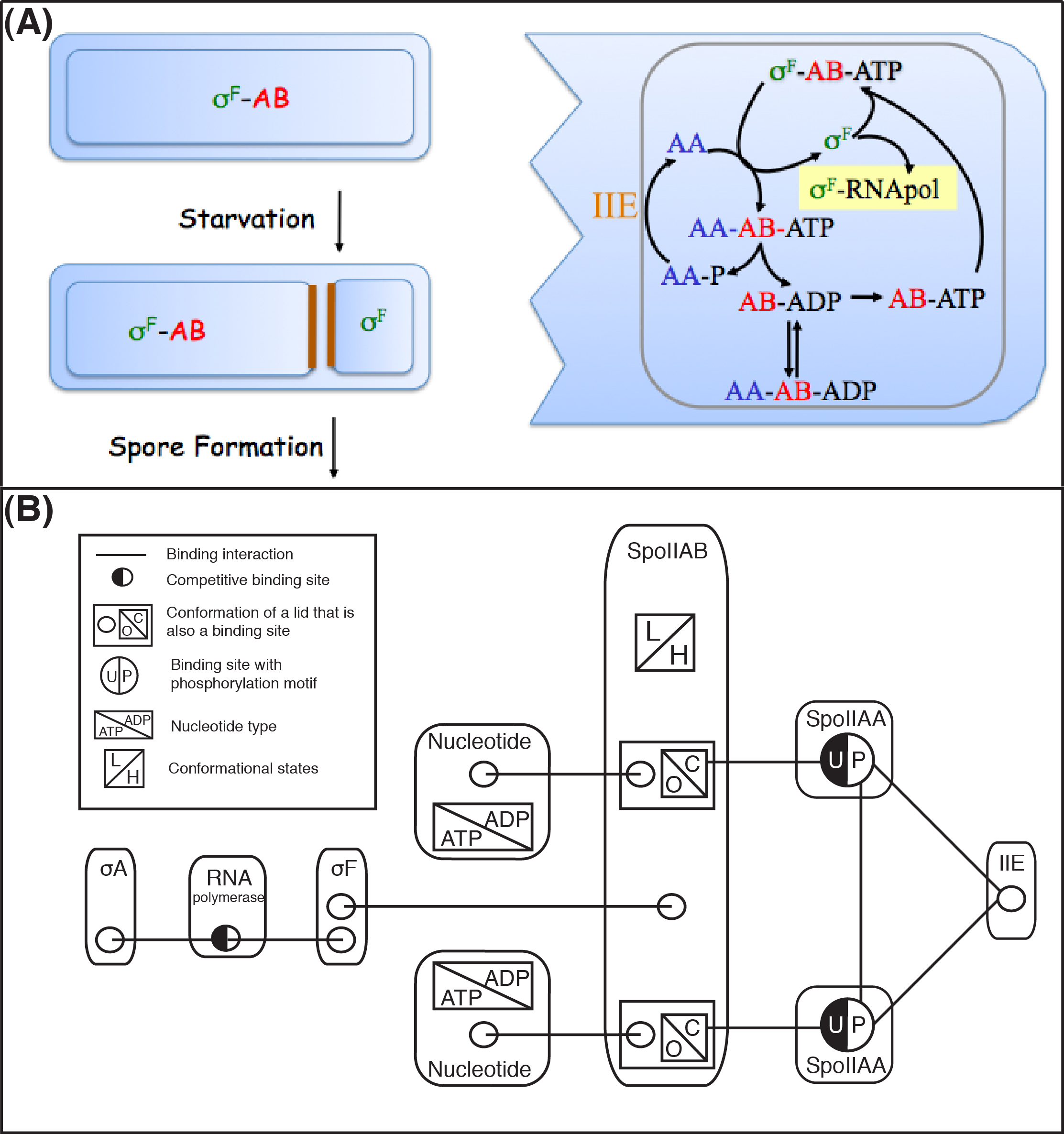}
\end{center}
\caption{{\bf Graphical representation of the model.} \textbf{(A)} (Left) Cartoon, illustrating the asymmetric division of bacteria as response to starvation, and the formation of a septum between the larger mother cell from the pre-spore. (Right) Cartoon of the biochemical interactions regulating $\sigma^F$ during sporulation in \textit{B. Subtilis}. The activity of $\sigma^F$ is controlled by the proteins SpoIIAB (AB), SpoIIAA (AA), and SpoIIE (IIE). AB binds $\sigma^F$ and keeps it in an inactive state. Binding of AA to the complex triggers the release of $\sigma^F$ which is then free to bind the RNA polymerase. This figure has been reproduced from a previous publication \cite{iber2006mechanism}.  \textbf{(B)} Contact map according to the rule-based formulation. The molecules involved in the regulation are indicated with smoothened rectangular shapes. As shown in the legend, interactions between molecules are indicated with line segments connecting their corresponding binding sites, which are indicated with circles. Competitive binding sites are half black. Rectangular shapes on the molecules indicate conformational changes, including closed/open, High/Low affinity, and nucleotide types. The circles that are both black and white indicate competitive binding sites. The rules in the Methods part specify these interactions.}
\label{Fig1}
\end{figure}

In a first step, we sought to formulate a comprehensive ODE model for the regulatory network. The previously published model had been hand-written and followed the interactions of 4 components ($\sigma^F$, AB, AA, IIE) and their modified forms and complexes to describe the \textit{in vitro} experiments \cite{iber2006mechanism}. To describe the \textit{in vivo} situation 2 further components ($\sigma^A$ and RNA polymerase) had been added in the original model. Given the combinatorical complexity (compare Fig. \ref{Fig1}A and Fig. S1), the original model comprised 50 ODEs to describe 150 different reactions between these few components, even though a number of less relevant reactions and states were ignored. Rule-based models facilitate the formulation of differential equation based models when multiple interactions among the components gives rise to combinatorial complexity and thus allows the efficient formulation of comprehensive models. We therefore formulated a rule-based model as graphically summarized in Fig. \ref{Fig1}B \cite{hlavacek2006rules}. The rule-based model was formulated to be applicable to both the \textit{in vitro} and \textit{in vivo} situations, and it also considered nucleotides as a variable species. Accordingly, the rule-based model included  seven molecule types. The network was described by 12 main rules (as specified in the Methods section). In brief, the activity of the Master transcription factor $\sigma^F$ is controlled by three proteins: SpoIIAB (AB), SpoIIAA (AA), and SpoIIE (IIE). AB binds $\sigma^F$ and keeps it in an inactive state. Binding of AA to the complex triggers the release of $\sigma^F$ which is then free to bind the core RNA polymerase, thereby forming the active holoenzyme, which can be directed to the transcriptional sides. AB acts as a kinase and phosphorylates bound AA, and IIE is the corresponding phosphatase de-phosphorylating AA. As part of the phosphorylation reaction, one ATP is converted into ADP at the nucleotide-binding site.  We showed previously that the AB dimer binds AA cooperatively, thus enhancing the sensitivity of the mechanism \cite{iber2006mechanism}. Another transcription factor, $\sigma^A$, is competing with $\sigma^F$ for RNA polymerase binding. The rules were translated into a set of 59 state variables and 190 individual reactions using BioNetGenerator \cite{faeder2009rule}. All the reactions were assumed to take place in the cytoplasmic compartment. The system was solved using the numerical integrator ode15s in MATLAB.

\subsection*{Parameter screens to identify parameter ensembles}

\begin{figure}[!ht]
\begin{center}
\includegraphics[height=15cm]{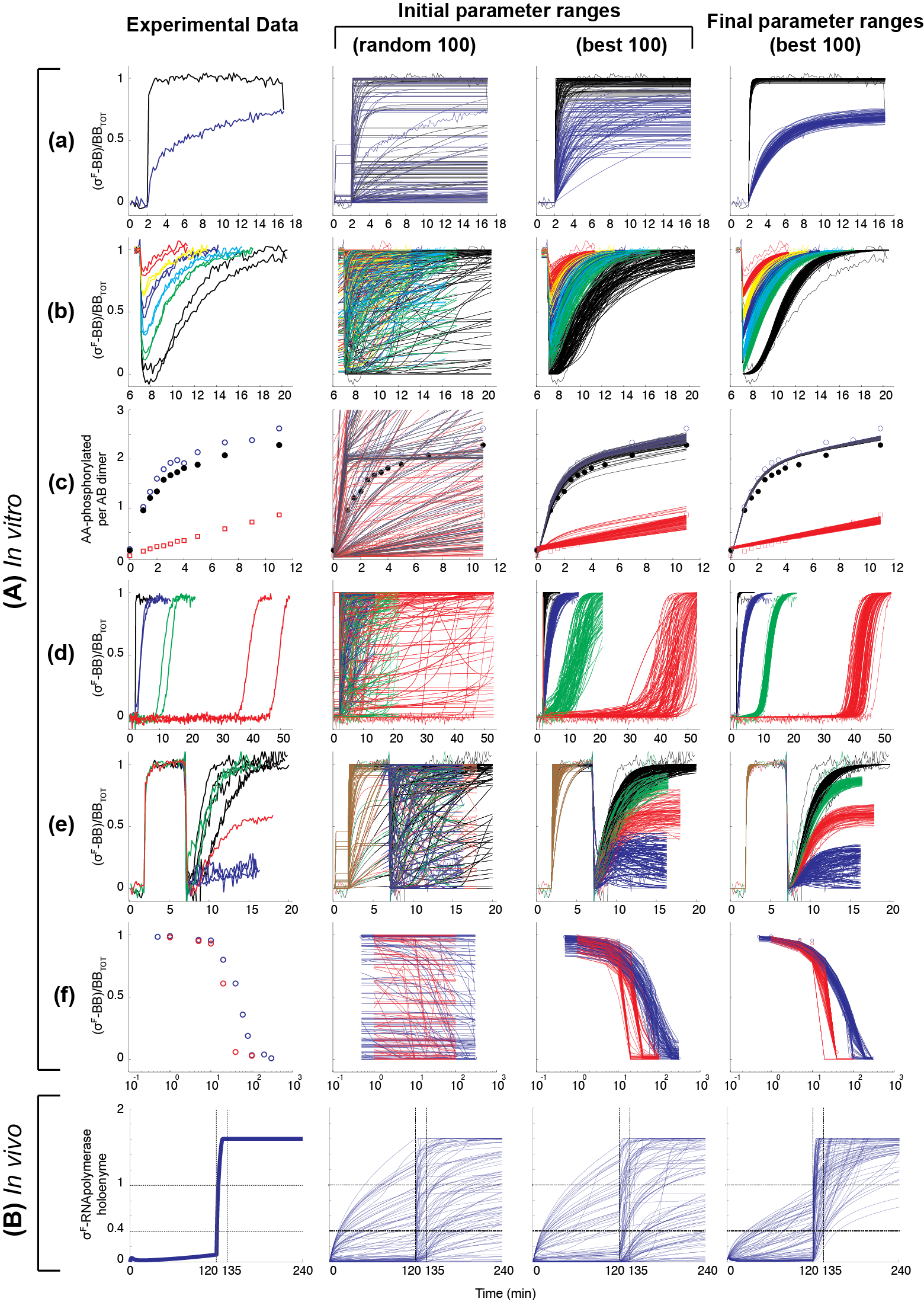}
\end{center}
\caption{{\bf Comparison of Model Predictions and Experimental Data.} (A) Six panels of time-resolved \textit{in vitro}  experimental data, measuring the fraction of $\sigma^F$-bound AB (a,b,d-f), and the ratio of phosphorylated AA per AB dimer (c). Different colors within the same panel indicate different experimental conditions as described in detail in \cite{iber2006mechanism}; for details see the method section. $10^6$ simulations of these \textit{in vitro} experiments were performed using the initially defined parameter ranges. 100 randomly selected model outputs, and 100 best ranked according to their sum squared residual errors are illustrated. (B) Behavior of the \textit{in vivo} response, according to \cite{iber2006mechanism}.  Using the same parameters of the \textit{in vitro} above, we simulate the \textit{in vivo} conditions, observing the formation of the $\sigma^F$-RNApolymerase holoenzyme during the sporulation process.}
\label{Fig3}
\end{figure}

Experimental data exist for both the  \textit{in vitro} and \textit{in vivo} situation. The \textit{in vitro} experiments characterized the interactions between AB, AA, IIE and $\sigma^F$ in the absence of $\sigma^A$ and core RNA polymerase (Fig.\ref{Fig3}A, first column). The first panel of Figure \ref{Fig3}B shows the key read-out of the physiological response, the concentration of the $\sigma^F$-bound RNA polymerase holoenzyme. To support sporulation, the concentration of $\sigma^F$-bound RNA polymerase holoenzyme must be negligible before septation, and must subsequently reach micromolar concentrations within 15 minutes or less, as illustrated in the first panel of Figure \ref{Fig3}B. 

\begin{figure}[!ht]
\begin{center}
\includegraphics[height=15cm]{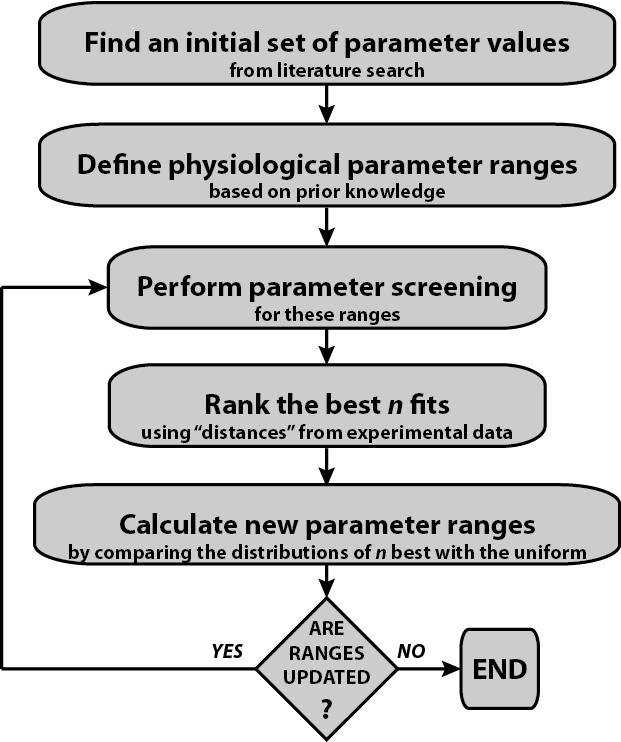}
\end{center}
\caption{{\bf Flowchart of the evolutionary sampling.} The initial parameter ranges were based on information available in the literature (Table 1). The ODE system was subsequently solved for multiple parameter sets sampled from these ranges. The sum of the squared residuals between the model predictions and the data was used to rank the parameter sets. Based on the best-\textit{n} fits the next parameter ranges were calculated. If the parameter ranges were updated the system was again simulated for the new parameter ranges. This process was followed until the parameter ranges could no longer be further updated.}
\label{Fig2}
\end{figure}

To run the simulations we had to define initial ranges for the parameter values.  Plausible ranges for the different parameter values are known from the literature. In previous similar sampling strategies, the parameter ranges were further adjusted by centering them around measured values from the literature  \cite{beyer2002evolution}. We followed a similar strategy and centered all parameter values around the previously estimated parameter values (Table \ref{tab1}). We then sampled from a range that extended to ten-fold higher and lower values for protein concentrations, and to 100-fold higher and lower values for reaction rates. We sampled $10^6$ parameter sets that were drawn at random from a log-uniform distribution. New parameter ranges for subsequent screens were defined based on the subset of the parameters corresponding to the best $10^3$ fits according to a least squared residual ranking (Fig. \ref{Fig2}). After multiple cycles of this process, the model exhibited stable model predictions for the \textit{in vitro} behavior. The visual results of this iterative process in terms of the model behavior are illustrated in Fig.\ref{Fig3}A.

Time resolved data of six different experimental settings monitor the fraction of $\sigma^F$-bound AB (\ref{Fig3}A,a,b,d-f), and the ratio of phosphorylated AA per AB dimer (\ref{Fig3}A,c)  under different conditions, as described in detail previously \cite{iber2006mechanism}. Each experiment focuses on characterizing different aspects of the system:  association of nucleotides with AB (\ref{Fig3}A,a,c,d), binding of AA with AB (\ref{Fig3}A,b,d,f), phosphorylation of AA (\ref{Fig3}A,b), and the effect of the phosphatase IIE in the dissociation of the $\sigma^F$-AB complex (\ref{Fig3}A,e,f). Considering the normalized contribution of each of these six experiments to the overall sum of residuals of the system, we can rank them according to their fitness. However, the sum of the square residuals, as a measure of fitness, does not suffice to capture behaviors, which would be considered important, such as the steepness of the response curves in \ref{Fig3}A,d.  Therefore, in addition to the squared residuals, the steepness of the response was also taken into account for this experiment. As expected, already in the initial parameter sets, there is a visible difference in the quality of 100 random fits compared to the 100 best fits. Nonetheless, it is only after several screens that the 100 best parameter sets allow a close fit of the model to the experimental data (Fig. \ref{Fig3}A).

\begin{figure}[!ht]
\centering
\includegraphics[width=11cm]{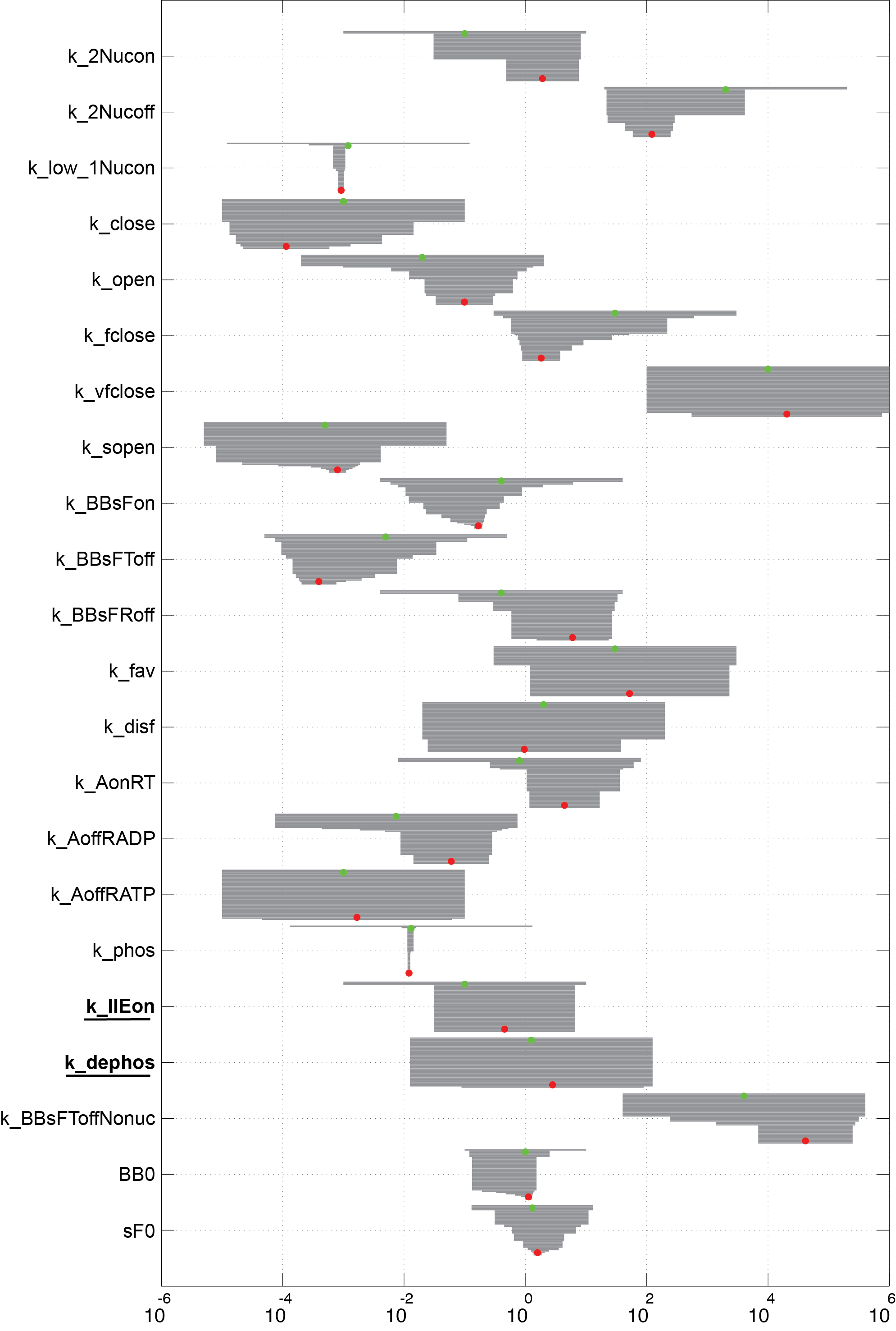}
\caption{{\bf The evolution of parameter ranges.} All parameters, which became constrained during the screening, are listed. The reference value is indicated by green dots, indicating the centers of the initially sampled parameter ranges. These ranges extended to ten-fold higher and lower values for protein concentrations, and to 100-fold higher and lower values for reaction rates. The sequential restriction of parameter ranges during the screening is illustrated with gray shadow. Red dots indicate the central values of the final parameter ranges. The parameters k\_IIEon and k\_dephos (bold underlined) are two of the parameters, which needed to be further restricted to reproduce the \textit{in vivo} behavior (\fig{Fig5}, Table 2).}
  \label{Fig4}
\end{figure}

The sequential restriction of parameter ranges during the iterations of the screen is shown in Fig. \ref{Fig4}, and is summarized in the last column of Table 2. During this evolution of parameter ranges only few parameter ranges become strongly restricted (Fig. \ref{Fig4}, Table 2). Among these are the initial concentrations of both AB and $\sigma^F$. The initial concentrations of the other components directly depend on those of AB and $\sigma^F$, and are therefore equally restricted. The protein concentrations were adjusted carefully in the \textit{in vitro} experiments, and while absolute concentration measurements are always difficult, the relative concentrations could be adjusted well, and we therefore fixed the relative concentrations also in the model. The fact that the screens greatly restricted the initial concentrations of both AB and $\sigma^F$ confirms the high sensitivity of the network to the concentrations of the players. 

Among the kinetic rates, the ranges of the  phosphorylation rate of AA, the rate of switching from ADP-bound to ATP-bound  AB, the rates of complex association and dissociation of BB with $\sigma^F$, the opening and closing rates of the AB lids covering the pockets of the nucleotide binding sites, and the rates of nucleotide association and dissociation with AB were strongly restricted (Fig. \ref{Fig4}). The other kinetic ranges that were affected by the screens  were less constrained, and 7 parameters were not affected at all by the screen (Fig. \ref{Fig4}). 

The differences in the extent, to which parameter ranges become restricted, reflect the type of data that was available. Thus the phosphorylation data in Fig. \ref{Fig3}A c strongly restrict the phosphorylation rate. On the other hand, the four binding rates of the RNA polymerase are not restricted at all by the \textit{in vitro} data as the RNA polymerase was not included in the experiments. Similarly, the SpoIIE-dependent rates are only partially restricted as there is only a single experiment (Fig. \ref{Fig3}A e,f) that involves SpoIIE. The SpoIIAA-P dimerization rates and the SpoIIE-AAp off-rate are not at all constrained. Finally, not all of the AB-AA rates have become constrained, but at the same time the screen did not reproduce the small difference between the black and the blue datasets in Figure \ref{Fig3}A, c, which is obtained when AB is pre-incubated for 5 minutes with ADP before ATP and AA are added (black dataset).

\subsection*{Prediction of physiological behavior}

The iterative screening of the parameters was based only on the best fits of the \textit{in vitro} experiments (Fig. \ref{Fig3}A). We wondered in how far the optimization of the parameters with the \textit{in vitro} data would improve their predictions of the physiological behaviour. The physiological response is mediated by $\sigma^F$-bound RNA polymerase holoenzyme. To support sporulation, $\sigma^F$-bound RNA polymerase holoenzyme must be negligible before septation, and must subsequently reach micromolar concentrations within 15 minutes or less, as illustrated in the first panel of Figure \ref{Fig3}B. Accordingly, we analyzed the time evolution of the  $\sigma^F$-RNA-polymerase holoenzyme for the different parameter sets. 

\begin{figure}[!ht]
\begin{center}
\includegraphics[height=5cm]{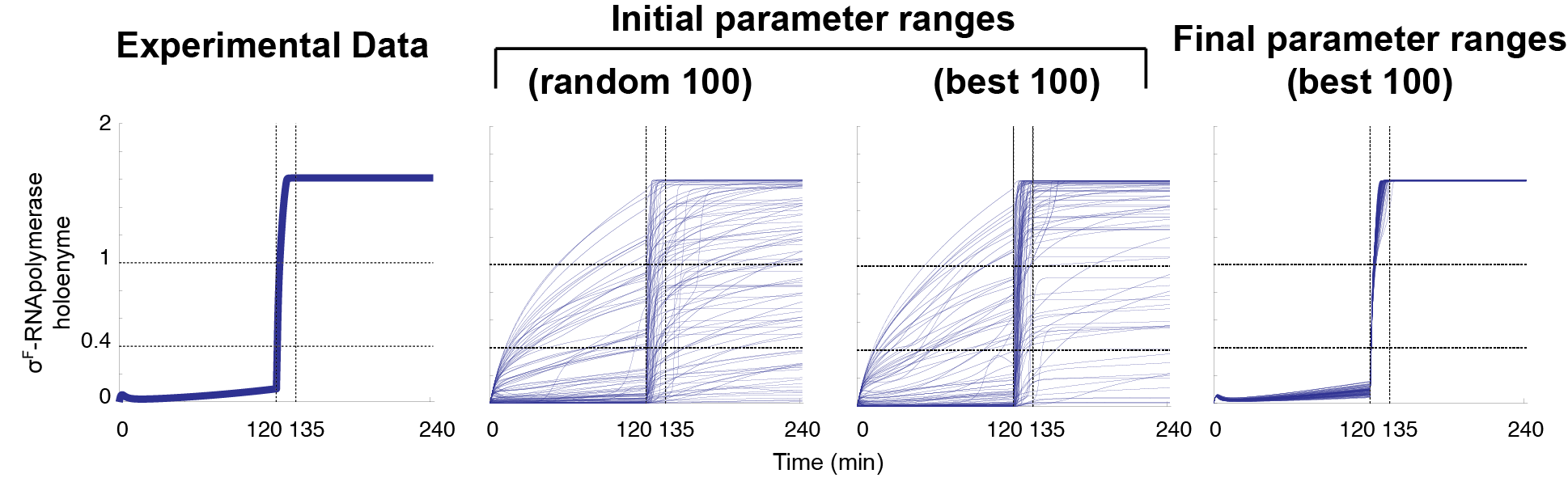}
\end{center}
  \caption{{\bf Predictive value of the optimized parameter sets for the physiological behavior.} The panels are the same as in \fig{Fig3}(B), but simulated after fixing further parameter values based on data in \cite{iber2006mechanism}, i.e. the AA dephosphorylation rate ($k\_dephos = 1.26 s^{-1}$), the AA-p IIE on \& off rates ($k\_IIEon = 9\times10^{-2} \mu M^{-1}s^{-1}$, $k\_IIEoff = 5.8\times10^{-1} \mu M^{-1}s^{-1}$), and the AAp dimer on \& off rates ($k\_dimon = 2\times10^{-1} \mu M^{-1}s^{-1}$, $k\_dimoff = 1s^{-1}$).}
\label{Fig5}
\end{figure}

As can be seen in Figure \ref{Fig3}B, the fraction of successful physiological responses increases as the parameter values are optimized for  \textit{in vitro} conditions. Many simulations, however, still fail. By comparing the distribution of each individual parameter in the successful the unsuccessful \textit{in vivo} simulations of the final screen (Fig.\ref{Fig5}, Table S2), we identified the critical parameters as the ones associated with the IIE-dependent dephosporylation of AA-P, i.e. the dephosphorylation rate and the association/ dissociation rates of the phosphatase (IIE) with its substrate AA-P, as well as the rates of AA-P dimerization. A comparison of the final ranges to the initial ranges of these parameter values (Fig.\ref{Fig5}, Table S2)  shows that these rates had not been restricted by the \textit{in vitro} experiments in Fig. \ref{Fig3}A. We therefore needed additional data to further restrict these. The previous study identified kinetic rates for these parameter values, based on further NMR experiments \cite{iber2006mechanism}. Once fixed to these values,  all simulations predicted the physiological data correctly (Fig. \ref{Fig5}), even though still most parameters remained poorly restricted (Fig. \ref{Fig4}).

\section*{Discussion}
 
 Mathematical models are increasingly used to define mechanisms in biology, but typically insufficient data is available to determine all parameter values with high confidence. Most mathematical models have previously only been analysed for one optimal parameter set. Recent advances in computing power now allow the screening of larger parameter spaces. We have exploited this to reanalyse a  previously published model, whose predictions had been confirmed in experiments. The model describes the regulation of $\sigma^F$ during sporulation of \textit{Bacillus subtilis}  \cite{iber2006mechanism}. By sampling from the parameter ranges we were able to fit the detailed experimental data, while imposing restrictions on only some of the parameter ranges. Subsequently we tested in how far the parameter sets, that allowed us to reproduce the measured \textit{in vitro} data, would also reproduce also the \textit{in vivo} behavior. We found that although the \textit{in vivo} behavior improved, still many of the parameter sets failed to reproduce the \textit{in vivo} behavior. By comparing successful and unsuccessful simulations we identified the critical parameter values and fixed these with further data. This let to the restriction of further parameter ranges, and allowed us to reproduce the physiological behavior. The latter shows that the parameterized model can also be used for experimental design, as it defines the critical rate constants that need to be measured to understand a biological behaviour of interest.
 
Most parameter ranges remain unconstrained, a phenomenon known as parameter sloppiness. Many models in system biology exhibit parameter sloppiness \cite{gutenkunst2007universally}. In fact, there may be an evolutionary role for parameter sloppiness as systems with sloppy parameters are more robust to changes. Systems with sloppy parameters could thus be optimized for various distinct functions without hampering other functions \cite{Soyer:2006p11564}. Parameter sloppiness also does not necessarily reduce the reliability of the model predictions, at least as long as the unconstrained parameters do not impact on the model predictions. Thus if both the experimental data and the biological question of interest do not depend on a particular parameter value then it is not a problem if this value is not constrained. However, typically it is very difficult to evaluate such aspect. Local sensitivity analyses are only moderately helpful, as the predictions may extend beyond the local reach of such approximation. The advantage of using the evolutionary sampling approach outlined in this study is that it provides a broader view on the dynamic behaviour of the model that cannot be obtained by a local sensitivity analysis. 

We note, however, that in spite of being restricted to a single parameter set, the previous study arrived at predictions that were subsequently confirmed in experiments. This demonstrates that a carefully parameterized model can still lead to reliable predictions, even when studied only locally. In fact, there are ample examples of useful predictions and novel insight that could be gained, although many of the parameter values could not be firmly defined \cite{bailey2001complex}. In conclusion, despite the limits on the availability of quantitative data, mathematical models can still be very useful in providing novel insight and making interesting predictions.

\section*{Methods}

\subsection*{Rule description}

The following mechanistic rules summarize the well characterized biochemical interactions of the $\sigma^F$ system:

\begin{enumerate}
\item Nucleotides binding to AB (ADP to ATP)

The protein AB has two binding pockets for nucleotides (ATP, ADP). The nucleotide-binding sites are each covered by a flexible lid that can either be in an open or closed conformation. AB is found in two conformations with either low or high affinity for AA. High affinity AB with two open lids can bind two nucleotides (ATP, ADP) in its pockets, irrespective of whether AB is already AA or $\sigma^F$-bound. A single ADP can be exchanged for ATP whenever there is one site that is not AA-bound, irrespective of the AB conformation.

\item  Lid opening-closing

Protein-unbound AB that contains nucleotides in its pockets can open and close its two lids (simultaneously). Whenever AB is bound by either $\sigma^F$, or AA, or both, its lids close faster and the reverse reaction is then very slow.

\item AB-$\sigma^F$ interaction  ($\sigma^F$ inactivation)

$\sigma^F$ can bind to AB as long as it has no more than one AA already bound. This interaction does not depend on its lid state or its conformation, but the affinity depends on the nucleotides bound to AB, i.e. the affinity is higher when AB has at least one ATP in its pockets. If AB is nucleotide-free, then the affinity is very low.

\item AB is an allosteric enzyme

AB can assume either of two conformational states. The conformation of unbound and $\sigma^F$-bound AB is biased towards a conformation that binds AA with low affinity. AA biases the conformational equilibrium of AB to the high affinity conformation.

\item AB-AA interaction

AB can bind to AA as long as $\sigma^F$ and another AA are not already both bound to it. The pockets of AB that are ATP-bound have a higher affinity for AA than those that are ADP-bound. Nucleotide-free AB can still bind AA, albeit with much lower affinity.

\item Exchange between ADP and ATP

The total amount of nucleotides is conserved throughout each experiment. In the \textit{in vitro} case the total concentration of nucleotides is fixed; therefore this reaction is not modeled explicitly. For the \textit{in vivo} case, this is considered to be a relatively fast equilibrium, keeping the ratio of ATP to ADP fixed.

\item AA phosphorylation

Whenever AA is in a complex with AB and it is bound at the ATP-bound pocket, it can dissociate upon phosphorylation, leaving AB with an ADP at that pocket.

\item AA de-phosphorylation

IIE phosphatase binds to phosphorylated AA (AA-p) and catalyses the release of the phosphate group.

\item AA-p dimerization

Free AA-p can form a homo-dimer complex.

\item Competition of  $\sigma^F$ and $\sigma^A$ for the RNA-polymerase

Both $\sigma^F$ and $\sigma^A$ form a complex with the core RNA polymerase. Binding of $\sigma^A$ is inhibiting the formation of the $\sigma^F$-holoenzyme.

\item Production (\textit{in vivo})

AB, AA, $\sigma^F$ and IIE can all be produced, while  the concentrations of the core RNA polymerase and of $\sigma^A$ are considered to be constant on the timescales of these experiments.
 
\item Degradation (\textit{in vivo})

AB is degraded only when unbound. The other proteins have an effective production rate that compensates for their degradation. 

\end{enumerate}

\subsection*{Differences between \textit{in vitro} and \textit{in vivo} conditions}

In the \textit{in vitro} experiments, core RNA polymerase and $\sigma^A$ are not  present, and these concentrations are therefore set to zero in the simulations. From the previous publication we have data for six distinct experimental set-ups that make use of fluorescence quenching measurements and measurement of protein phosphorylation (Fig \ref{Fig3}A). The experiments have been described in detail previously \cite{iber2006mechanism}. In brief, Figure \ref{Fig3}A(a) reports the fraction of $\sigma^F$ (total 1.3 $\mu$M) that is bound in a complex with AB dimer (total 1 $\mu$M) over time in response to the addition of 100 $\mu$M ATP (black curve) or ADP (blue curve) after 2 minutes of co-culture in the absence of nucleotides. Figure \ref{Fig3}A(b) reports the kinetics of $\sigma^F$-AB unbinding and rebinding upon addition of AA (0.4 $\mu$M  (red),  0.7 $\mu$M (yellow), 1 $\mu$M (blue), 1.5 $\mu$M (cyan), 2 $\mu$M (green) and 3 $\mu$M (black)). Figure \ref{Fig3}A(c) reports the phosphorylation of AA per AB dimer when AB is directly incubated with 40 $\mu$M AA and 100 $\mu$M ATP (blue) or pre-incubated for 5 min with either 5 $\mu$M ADP (black) or 5 $\mu$M ADP and 40 $\mu$M AA (red). Figure \ref{Fig3}A(d) reports the fraction of $\sigma^F$ (total 1.3 $\mu$M) that is bound in a complex with AB dimer (total 1 $\mu$M) over time in the presence of different concentrations of AA (no AA (black), 2 $\mu$M (blue), 4 $\mu$M (green) or 6 $\mu$M (red) AA). Figure \ref{Fig3}A(e) reports the fraction of $\sigma^F$ (total 1.3 $\mu$M) that is bound in a complex with AB dimer (total 1 $\mu$M) over time. $\sigma^F$ and AB bind upon addition of ATP (2 min), and subsequently dissociate upon addition of 2.5 $\mu$M AA. The extent of $\sigma^F$-AB complex re-formation depends on the concentration of IIE (none (black), 10nM (green), 40nM (red) or 100nM). The last panel, Figure \ref{Fig3}A(f), reports the final fraction of $\sigma^F$-bound AB at different concentrations of IIE at two different concentrations of AA (2.5 $\mu$M (blue) or 4 $\mu$M (red)). 31 parameters were sampled (Table \ref{tab1}), two of them describing the initial concentrations (AB0 and sF0). In the bacterium, the genes that encode AB, AA, and $\sigma^F$ are localised on an operon, and the ratio between these concentrations is therefore strongly constrained \cite{iber2006quantitative}. Therefore, the concentration of AA is scaled with respect to the value of the sampled AB. The gene for IIE is not on this operon, and the IIE concentration is therefore fixed independently; its concentration can thus differ relative to the others within the sampled parameter space.

\textit{In vivo}, proteins can also be produced and degraded, and production and decay processes therefore need to be considered to model the \textit{in vivo} situation.  Experiments showed that the rate of de-phosporylation by IIE is 144-fold lower for the physiological salt concentrations than for the salt conditions typically used in \textit{in vitro} experiments \cite{iber2006mechanism}, and the dephosporylation rate was adjusted accordingly. Additionally, because of the smaller size of the prespore compartment relative to the mother cell and the accumulation of IIE on both sites of the septum (Fig.\ref{Fig1}), the effective concentration of IIE (and of the IIE-bound AA) increases 4-fold upon septation. To compare the two experimental conditions, we used the same parameter sets for the  \textit{in vitro} and the \textit{in vivo} simulations. The parameters that describe production/degradation that are absent in the \textit{in vitro} simualtions, are scaled based on the respective sampled values of AB0. Similarly, the concentration ranges of $\sigma^A$ and core RNA-polymerase were based on the parameter ranges of AB0. The degradation rate of AB was adjusted such that the ratio between degradation and production remains constant.

\subsection*{Parameterization and treatment of uncertainty} 

Previously estimated parameters of the system \cite{iber2006mechanism} were used as reference values that defined the center of rather wide sampling ranges (Table 1). The parameters describing reaction rates were sampled over four orders of magnitude, while the parameters that describe concentration were sampled over  two orders of magnitude. The parameters corresponding to the initial protein concentration levels were sampled over a narrower range, because their measurement is direct and therefore less prone to estimation errors. All parameters were sampled uniformly on a $\log_{10}$ scale; this way sampling is equally likely for any order of magnitude.

\subsection*{Optimizing the parameter sets} 

Starting with the initial parameter ranges, we sampled $10^6$ ($N_S$) parameter sets, integrated the system of ODEs, and compared the output to the experimental data (Fig. \ref{Fig3}A). We determined the sum of squared residuals (SSR) between the  simulated observable $f(x_i)$ and the experimental values $y_i$. So for each experiment

\begin{equation}
SSR_{k} = \sum_{i=1}^{n_k}  \sum_{j=1}^{|n_{k,i}|} (y_i - f(x_i))^2,  
\end{equation}

\noindent where $n_k$ represents the number of multiple curves within the same experiment, $n_{k,i}$ corresponds to the set of time points for the domain of each curve $n_k$ of the same experiment $k$, and can be different for each experiment. We need to define a measure for the overall residual that corresponds to each individual sampled parameter set $S$, given those differences between and within the different experiments. Therefore, following a naive but objective approach, we normalize each $SSR_{k}$ with respect to its corresponding size of data points. The measure of the total normalized residual then treats all experiments equally and is obtained as:

\begin{equation}
SSR_k^{s} = \sum_{k=1}^4 \frac{SSR_{k}}{\sum_i^{n_k} |{n_{k,i}}|},  %
\end{equation}

\noindent where $s$ is the index of the sampled parameter set. The contribution of noise to the data variation was not considered, because replicates of the same experiments show that the variance of the data does not scale with measurement size, and it is similar for the different experiments.  

Subsequently, we ranked the parameter sets according to the increasing order of this measure (e.g. the first parameter set corresponds to the best fit to the data), and investigated the effect of this ranking on each parameter. Therefore we compared different subset sizes of every ranked parameter with its corresponding total pool ($10^6$) to observe deviations from the initially assumed uniform distribution. The Kolmogorov-Smirnov test statistic was used as a measure to quantify the distance between these two distributions, i.e. the subset versus the whole pool ($10^6$) of each parameter. The p-value of the Kolmogorov-Smirnov test had a minimum for comparisons with subsets of size $10^3$. Thus this subset size was used as representative to indicate affected parameters as imposed by the data. This statistic is useful to highlight the distance between distributions, but it is not informative regarding the qualitative differences of these distributions. For this reason we define here a heuristic cut-off criterion to distinguish whose parameters subset distributions impose different ranges. This heuristic compares the initial range of each parameter with the range that is defined by the middle $90\% $ of its $10^3$ ranked subset. So, for parameter $\kappa$, if

\begin{equation}
\frac{Q^{\kappa}_{.95}- Q^{\kappa}_{.05}}{range(\kappa)}<0.8
\end{equation}

\noindent then we further restrict the initial parameter range. To enhance the estimation quality of these quantiles, also in the case of very flat tails, additional bootstrapping was performed for the lower bound ($5\%$) of the $5\%$ quantile and of higher bound ($95\%$) of the $95\%$ quantile, and the criterion was then tested for the $Q^{Q^{\kappa}_{.05}}_{.05}$ and the $Q^{Q^{\kappa}_{.95}}_{.95}$. The above quantiles define a sub-domain of the initial parameter ranges, and the ranges of the parameters that meet this criterion are then restricted to their corresponding sub-domain. This process is repeated as long as the criterion is met for at least one parameter Fig. \ref{Fig2}.

\subsection*{Definition of physiological response}

The \textit{in vivo} experiment starts at time zero with production of all species apart from $\sigma^A$ and RNA-polymerase that are considered to be constant for the timescale of the experiment. At $t=2h$ septation happens. Upon septation, the production of the proteins stops, and the effective concentration of IIE (and IIE-bound AA) is increased by four-fold, because of the difference in cell size. The increase in IIE and IIE-AA levels is immediate, but the response of the $\sigma^F$ holoenzyme formation takes about 15min. For the response to be physiological, the level of holoenzyme before septation must not exceed 0.4 $\mu M$, otherwise we would have septation-independent holoenzyme formation. The holoenzyme should be above $1\mu M$ 15 min after the septation, and then still remain high for some time. Therefore a successfull physiological response must fulfill the following:

\begin{enumerate}
\item t = 2h : [$\sigma^F$-RNA-polymerase]  $<$ 0.4
\item t = 2h15min : [$\sigma^F$-RNA-polymerase] $>$ 1
\item t = 4h : 1$<$ [$\sigma^F$-RNA-polymerase]  $<$ 4
\end{enumerate}

\newpage

\bibliographystyle{naturemag}

\newpage 
\section*{Acknowledgements}
This research was supported by an iPhD SystemsX grant. The authors thank Florian Geier for valuable discussions.

\section*{Authors contribution statement}
GF adopted the published model by DI, GF developed the infrastructure for the parameter screens,  GF and DI analyzed the results and wrote the paper.    

\section*{Competing financial interests}
Herewith it is declared that the authors have no competing financial, professional or personal interests that might have influenced the performance or presentation of the work described in this manuscript.

\newpage 
\section*{Figures}

\newpage
\clearpage

\section*{Tables}

\begin{table}
\small
  \label{tab1}
  \renewcommand\arraystretch{1}
  \centering
  \begin{tabular*}{1\textwidth}{@{\extracolsep{\fill}}lllll}
    \hline
    \# & Parameter description & Parameters & Basal values & Sampled ranges\\
    \hline
    \underline{1}   & Nucleotide-AB on-rate & k\_2Nucon & $10^{-1} \mu M^{-2}  s^{-1}$ & $ [10^{-3}-10^{1}]$ \\
    \underline{2}   & Nucleotide-AB off-rate  & k\_2Nucoff  & $2\times 10^3 s^{-1}$ & $2\times [10^1-10^5]$ \\
    \underline{3}   & ATP-ADP exchange rate for AB-bound AA & k\_low\_1Nucon & $1.2\times10^{-3} \mu M^{-1}  s^{-1} $ & $1.2\times[10^{-5}-10^{-1}]$ \\
    \underline{4}   & AB-lid closure & k\_close & $10^{-3} s^{-1}$ & [$10^{-5}$-$10^{-1}$] \\
    \underline{5}   & AB-lid opening & k\_open & $2\times10^{-2} s^{-1}$ & $2\times[10^{-4}-10^{0}]$ \\
    \underline{6}   & ADP-bound AB-complex-lid closure & k\_fclose & $3\times10^{1} s^{-1}$ & $3\times[10^{-1}-10^{3}$] \\
    \underline{7}   & ATP-bound AB-complex-lid closure & k\_vfclose & $10^{4}  s^{-1} $ & [$10^{2}$-$10^{6}$] \\
    \underline{8}   & Nucleotide-bound complex-AB-lid closure & k\_sopen & $5\times10^{-4}  s^{-1} $ & $5\times[10^{-6}-10^{-2}]$ \\
    \underline{9}   & $\sigma^F$-AB on rate & k\_ABsFon & $0.4 \mu M^{-1}  s^{-1} $& $4\times[10^{-3}-10^1]$ \\
    \underline{10} & $\sigma^F$-AB$^L$(Low affinity for AA) on-rate & k\_ABsFToff & $5\times10^{-3} s^{-1}$ & $5\times[10^{-5}-10^{-1}]$ \\
    \underline{11} & $\sigma^F$-AB$^H$(High affinity for AA) on-rate & k\_ABsFRoff & $0.4 s^{-1}$ & $4\times[10^{-3}-10^1]$ \\
    \underline{12} & ADP-bound AB conformational change & k\_fav & $3\times 10^{1} s^{-1}$ & $3\times[10^{-1}-10^3]$ \\
    \underline{13} & ATP-bound AB conformational change & k\_fav\_hi & $5\times10^{4} s^{-1}$ & $5\times[10^{2}-10^{6}]$ \\
    \underline{14} & Rate of disfavored conformational change & k\_disf & $2 s^{-1} $& $2\times[10^{-2}-10^{2}]$ \\
    \underline{15} & AB-AA on-rate & k\_AonRT & $8\times10^{-1} \mu M^{-1}  s^{-1}$& $8\times[10^{-3}-10^1]$ \\
    \underline{16} & AB-AA off-rate & k\_AoffT & $2.4 s^{-1}$& $2.4\times[10^{-2}-10^{2}]$ \\
    \underline{17} & ADP-bound-AB$^H$ - AA off-rate & k\_AoffRADP & $7.4\times 10^{-3} s^{-1}$ & $7.4\times[10^{-5}-10^{-1}]$ \\
    \underline{18} & ATP-bound-AB$^H$ - AA off-rate & k\_AoffRATP & $10^{-3} s^{-1}$ & $[10^{-5}-10^{-1}]$ \\
    19 & ADP to ATP exchange& \textbf{k\_DT} &$10^4 s^{-1}$& fixed \\
    20 & ATP to ADP exchange & \textbf{k\_TD} &$10^3 s^{-1}$& fixed \\
    \underline{21} & AA phosphorylation rate & k\_phos & $1.3\times 10^{-2} s^{-1}$ & $1.3\times[10^{-4}-10^{0}]$ \\
    \underline{22} & AA-p - IIE on-rate & k\_IIEon & $10^{-1} \mu M^{-1}  s^{-1}$ & $[10^{-3}-10^{1}]$ \\
    \underline{23} & AA-p - IIE off-rate & k\_IIEoff & $5.8\times10^{-1}$ & $5.8\times[10^{-3}-10^{1}]$ \\
    \underline{24} & AA dephosphorylation rate & k\_dephos & $1.25 s^{-1}$& $1.25\times[10^{-3}-10^{1}]$ \\
    \underline{25} & $\sigma^{F/A}$-RNA polymerase on-rate & \textbf{k\_spolon} & $1 \mu M^{-1}  s^{-1}$ & $[10^{-2}-10^{2}]$ \\
    \underline{26} & $\sigma^F$-RNA polymerase off-rate & \textbf{k\_sFpoloff} & $5.5\times10^{-1} s^{-1}$ & $5.5\times[10^{-3}-10^{1}]$ \\
    \underline{27} & $\sigma^A$-RNA polymerase off-rate & \textbf{k\_sApoloff} & $2\times10^{-2} s^{-1} $ & $2\times[10^{-4}-10^{0}]$ \\
    \underline{28} & AA-p dimer on-rate & k\_dimon & $2\times10^{-1} \mu M^{-1}  s^{-1}$ & $2\times[10^{-3}-10^{1}]$ \\
    \underline{29} & AA-p dimer off-rate & k\_dimoff & $1 s^{-1}$ & $[10^{-2}-10^{2}]$ \\
    30 & AB degradation rate& \textbf{k\_degr\_AB} & $s^{-1}$ & scaled \\
    \underline{31} & Open-lid-AB$^L$ - AA off-rate & k\_AoffT\_opl & $8\times10^{2} s^{-1}$ & $8\times[10^{0}-10^{4}]$ \\
    \underline{32} & $\sigma^F$ - nucleotide-unbound-AB$^{L/H}$ off rate & k\_ABsFToffNonuc & $4\times10^{3} s^{-1}$ & $4\times[10^{1}-10^{5}]$ \\
    33 & AB synthesis rate & \textbf{k\_synth\_AB} & $6\times10^{-3}  \mu M  s^{-1}$ & scaled \\
    34 & AA synthesis rate & \textbf{k\_synth\_A} & $6\times10^{-3} \mu M  s^{-1}$ & scaled \\
    35 & $\sigma^F$ synthesis rate & \textbf{k\_synth\_sF} & $2\times10^{-3} \mu M  s^{-1}$ & scaled \\
    36 & IIE synthesis rate & \textbf{k\_synth\_IIE} & $2\times10^{-3} \mu M  s^{-1}$ & scaled \\
    37 & AA initial concentration & A0 & $\mu$M & scaled\\
    \underline{38} & AB initial concentration & AB0 & 1 $\mu M$ & $[10^{-1}-10^{1}]$ \\    
    \underline{39} & $\sigma^F$ initial concentration & sF0 & 1.3 $\mu M$ & $1.3\times[10^{-1}-10^{1}]$ \\
    40 & $\sigma^A$ initial concentration & sA0 & $\mu M$ & scaled \\
    41 & RNA polymerase initial concentration & \textbf{RNApol0} & $\mu M$ & scaled \\
    42 & ATP initial concentration & ATP0 & $\mu M$ & scaled \\
    43 & ADP initial concentration & ADP0 & $\mu M$ & scaled \\
    44 & IIE initial concentration & IIE0 & $\mu M$ & scaled \\
    \hline
  \end{tabular*}
    \caption{\textbf{Model parameters.} Parameter descriptions, names, basal values and units, and initially sampled ranges. The model parameters comprise kinetic rate constants (1-36), and protein concentrations (37-44). In the \textit{in vitro} case 31 parameters were sampled (underlined). The initial ranges of the kinetic parameters span four, and the initial concentrations two orders of magnitude. The parameters in bold were only used in the \textit{in vivo} case. Parameters, which were not sampled, either have fixed values, or they are scaled relative to other sampled parameters.}
\end{table}

\begin{table*}[h]
\small
\label{tab2}
\renewcommand\arraystretch{1}
  \centering

\begin{tabular*}{1\textwidth}{@{\extracolsep{\fill}}lllll}
\hline
Parameter description & Parameters & p-value (K-S Test) & p-value (T-test) & $\times$100\% constrained\\
\hline
 AA dephosphorylation rate & k\_dephos & $<10^{-10}$   & $<10^{-10}$ & 0.29\\
 AA-p - IIE on-rate & k\_IIEon   & --"--   & --"-- & 0.34\\
 AA-p - IIE off-rate & k\_IIEoff & --"--   & --"-- & 0 \\
 AA-p - dimer on-rate & k\_dimon & --"--   & --"-- & 0 \\
 AA-p - dimer off-rate & k\_dimoff & --"--   & --"-- & 0 \\
 ATP-bound AB conformational change & k\_fav & --"--   & --"-- & 0.23 \\
 Rate of disfavored conformational change & k\_disf   & --"--   & --"-- & 0.81\\
 AB-lid opening & k\_open   & --"--   & --"-- & 0.86\\
 ATP-bound-AB$^H$ - AA off-rate & k\_AoffRATP   & --"--   & --"-- & 0.38\\
 ATP-bound AB conformational change & k\_fav\_hi   & --"--  & --"-- & 0 \\
 $\sigma^F$-AB$^H$(High affinity for AA) on-rate & k\_ABsFRoff   & --"-- & --"-- & 0.45\\
 Open-lid-AB$^L$ - AA off-rate & k\_AoffT   & --"--  & --"-- & 0 \\
 ADP-bound-AB$^H$ - AA off-rate & k\_AoffRADP   & --"--  & --"-- & 0.67 \\
 $\sigma^F$-AB on rate & k\_ABsFon   & --"--  & --"-- & 0.99\\
 $\sigma^A$-RNA polymerase off-rate & k\_sApoloff   & 0.0932   & 0.1177 & 0 \\
Nucleotide-bound complex-AB-lid closure  & k\_sopen   & 0.1011   & 0.1816 & 0.98 \\
 AB initial concentration & AB0   &      0.1470   & 0.3253 & 0.96 \\
 $\sigma^{F/A}$-RNA polymerase on-rate & k\_spolon   & 0.1535   & 0.0510 & 0 \\
 ADP-bound AB-complex-lid closure & k\_fclose   & 0.2522   & 0.1873 & 0.99 \\
 AB-lid closure & k\_close   & 0.2559   & 0.0872 & 0.99 \\
 ATP-bound AB-complex-lid closure & k\_vfclose   & 0.2916   & 0.1700 & 0.24 \\
 AB-AA on-rate & k\_AonRT   & 0.3949   & 0.1497 & 0.80 \\
 Open-lid-AB$^L$ - AA off-rate & k\_AoffT\_opl   & 0.4110   & 0.5895 & 0 \\
 Nucleotide-AB off-rate & k\_2Nucoff   & 0.4663   & 0.2591 & 0.99 \\
 $\sigma^F$-RNA polymerase off-rate & k\_sFpoloff   & 0.5047   & 0.4955 & 0 \\
 AA phosphorylation rate & k\_phos   & 0.5364   & 0.2357 & 0.99 \\
 ATP-ADP exchange rate for AB-bound AA & k\_low\_1Nucon   & 0.5771   & 0.8398 & 0.99 \\
 $\sigma^F$ initial concentration & sF0   &      0.6507   & 0.9371 & 0.96 \\
 Nucleotide-AB on-rate & k\_2Nucon   & 0.6693   & 0.7025 & 0.28 \\
 $\sigma^F$ - nucleotide-unbound-AB$^{L/H}$ off rate & k\_ABsFToffNonuc   & 0.8480   & 0.8961 & 0.39 \\
 $\sigma^F$-AB$^L$(Low affinity for AA) on-rate & k\_ABsFToff   & 0.8898   & 0.5968 & 0.99 \\
\hline
\end{tabular*}
\caption{\textbf{Analysis of failure to reproduce the \textit{in vivo} behavior.} The sampled parameter sets of the last screen were divided into two groups, based on their success or failure to reproduce the \textit{in vivo} response. The equality of the distributions of each parameter in the two groups was tested using the non parametric Kolmogorov-Smirnov two-sample test. Similarly, the equality of the means of these distributions was tested with a two sample T-test. The p-values of these tests are listed. Comparing the initial and the final ranges, the last column indicates the percentage of constraint with respect to the initial parameter ranges.}
\end{table*}

\clearpage
\newpage

\section*{Supplementary Figures and Legends}

\renewcommand{\thefigure}{S\arabic{figure}}
\setcounter{figure}{0}

\begin{figure}[!ht]
\begin{center}
\includegraphics[width=18cm]{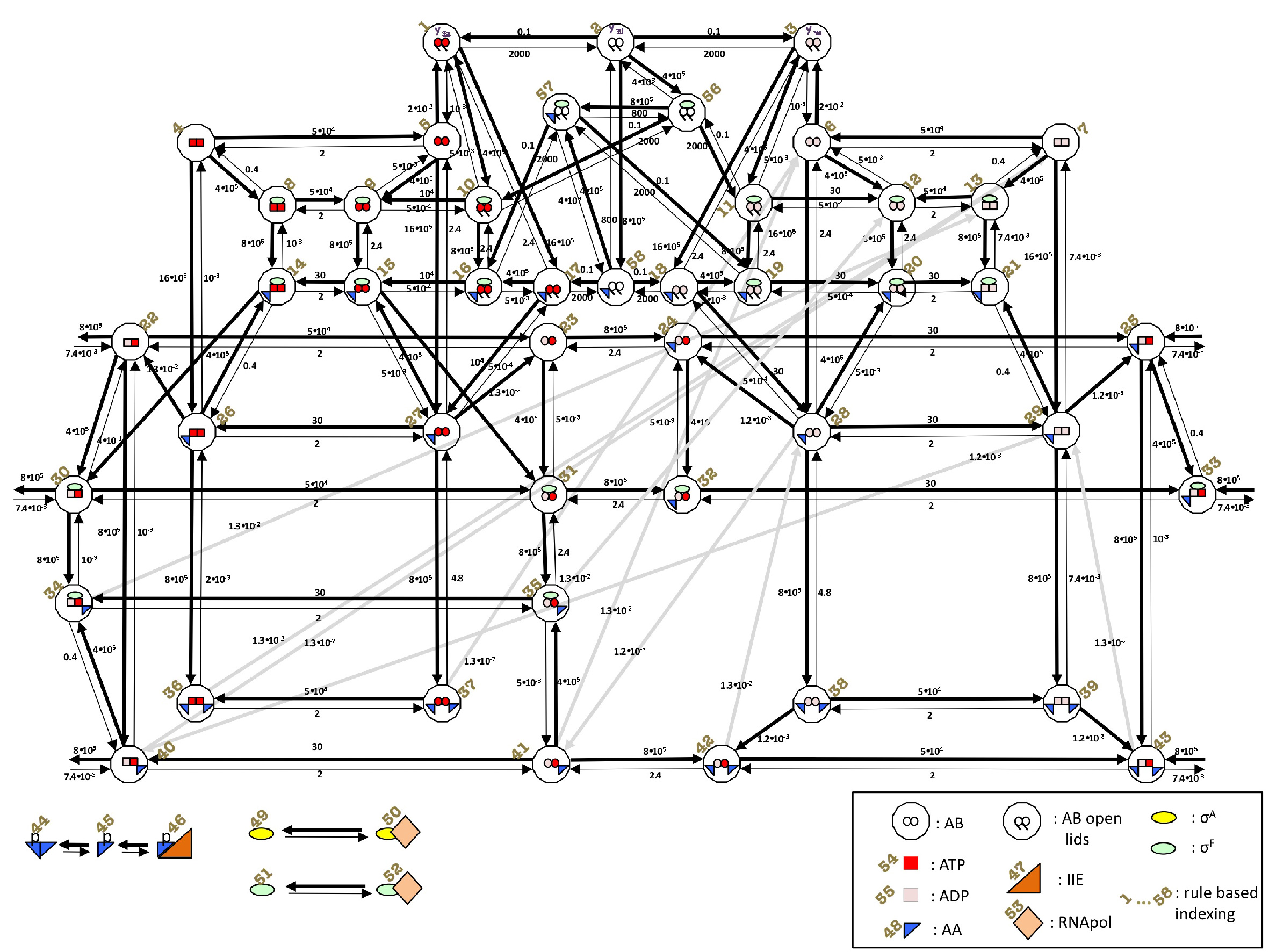}
\end{center}
\caption{{\bf Network of interactions.} Extended version of the already published model for $\sigma^F$. Detailed interactions of all species (in contrast to the compact representation of the contact map in the main text). The current scheme contains three additional species, in comparison to the already published model, which enable the interaction of the nucleotide-free BB with $\sigma^F$ and AA (i.e. 56,57,58), for slower reaction rates. Thicker arrows represent the faster reaction rates. Gray arrows are indifferent, just illustrating unidirectional reactions of species that are not close in the scheme.}
\label{supp2}
\end{figure}

\end{document}